\documentclass[fleqn,usenatbib,useAMS]{mnras}
\usepackage{graphicx}	% Including figure files
\usepackage{amsmath}	% Advanced maths commands
\usepackage{multicol}        % Multi-column entries in tables
\usepackage{bm}		% Bold maths symbols, including upright Greek
\usepackage{pdflscape}	% Landscape pages

\usepackage[T1]{fontenc}
\usepackage{ae,aecompl}

\usepackage{newtxtext,newtxmath}
\title[Transport-driven  Super-Jeans fragmentation]{{Transport-driven super-Jeans fragmentation in dynamical star-forming regions }}

\author[Guang-Xing Li]{Guang-Xing Li$^{1}$%
\thanks{Contact e-mail: \href{mailto:gxli@ynu.edu.cn}{gxli@ynu.edu.cn}, \href{mailto:ligx.ngc7293@gmail.com}{ligx.ngc7293@gmail.com}}%
\\
$^{1}$South-Western Institute For Astronomy Research, Yunnan University, Kunming, 650600, China}

\date{Last updated 2020 June 10; in original form 2013 September 5}

\pubyear{2020}

\begin{document}
\label{firstpage}
\pagerange{\pageref{firstpage}--\pageref{lastpage}}
\maketitle

\begin{abstract}
    The Jeans criterion is one cornerstone
    in our understanding of gravitational fragmentation. 
    A critical limitation of the Jeans criterion is that the background density
    is assumed to be a constant, which is often not true in
    dynamic conditions such as star-forming regions.  For example, during the formation phase of the high-density gas filaments in a molecular cloud, a density increase rate $\dot \rho$ implies an mass accumulation time of $t_{\rm acc}= \rho / \dot \rho= - \rho  (\nabla \cdot (\rho \vec{v}))^{-1}$. The system is non-stationary when the mass accumulation time becomes comparable to the free-fall time $t_{\rm
    ff} = 1 / \sqrt{G \rho}$.
    We study fragmentation in non-stationary
    settings, and find that accretion can
     significantly increases in the characteristic mass of
    gravitational fragmentation ( $\lambda_{\rm Jeans,\; aac}= \lambda_{\rm
    Jeans} (1 + t_{\rm ff} / t_{\rm acc})^{1/3}$,    
  $m_{\rm Jeans,\, acc}  = m_{\rm Jeans} (1 + t_{\rm ff} / t_{\rm acc})$). In massive star-forming regions, this mechanism of transport-driven super-jeans
  fragmentation can contribute to the formation of massive stars by causing
  order-of-magnitude increases in the mass of the fragments.
\end{abstract}


%%%%%%%%%%%%%%%%%%%%%%%%%%%%%%%%%%%%%%%%%%%%%%%%%%

%%%%%%%%%%%%%%%%% BODY OF PAPER %%%%%%%%%%%%%%%%%%

% The MNRAS class isn't designed to include a table of contents, but for this document one is useful.
% I therefore have to do some kludging to make it work without masses of blank space.
% In the standard picture, the fragmentation 
% occurs in a medium of density $\rho$ and velocity dispersion $\sigma_{\rm v}$.
% The crossing time is $t_{\rm cross} = \lambda / \sigma_{\rm v}$  and the
% free-fall time is $t_{\rm ff } = 1 / \sqrt{G \rho}$. The fragmentation length can be determined by letting 
% $t_{\rm cross} = t_{\rm ff}$, and the resulting length scale 

\section{Introduction }\label{sec1} 
\noindent 
The Universe is a gigantic multi-scale self-gravitating system. 
The Jeans criterion \citep{1902RSPTA.199....1J}  sets the foundation upon which our understanding of gravitational fragmentation bases.
$\lambda_{\rm Jeans} = \sigma_v  / (G \rho)^{1/2}$ is called the Jeans length, which is a threshold wavelength beyond which perturbations are gravitationally unstable. The Jeans criterion has a profound impact on modern astrophysical research as the core of the theory has been essential in understanding the behavior of a variety of astrophysical systems, including the fragmentation of gas under turbulent support \citep{2003ApJ...585..850M} and the stability of galaxy disks \citep{1964ApJ...139.1217T}.  The Jeans criterion is also a critical part of modern numerical simulations of self-gravitating gas as it has been integrated into the algorithm controlling the production of sink particles  \citep{2012MNRAS.419.3115B} -- representative particles created to remove gravity-included singularities in simulations.

One critical limitation of the Jeans criterion is that the initial density distribution is assumed to be static at the start. Whether the initial conditions are truly static has been a question barely addressed in previous studies. 
Contrary to the stationary appearance as often taken for granted,  most astrophysical systems are dynamic. This is particularly true for star-forming regions in the Milky Way, where density enhancements are generated by turbulence \citep{1998PhRvE..58.4501P,2011ApJ...730...40P}, converging flows \citep{1999A&A...351..309H,2007ApJ...657..870V} from turbulence \citep{2020ApJ...900...82P}, Global Hierarchical Collapse \citep{2019MNRAS.490.3061V} or cloud-cloud collisions  \citep{2021PASJ...73S...1F} in relatively short times. The rapid density increases imply another timescale, the mass accumulation time $t_{\rm acc} = \rho / \dot \rho$. When $t_{\rm acc}$ is shorter than the free-fall time, the effect of fast mass transport becomes significant.

We modify the Jeans formalism to extend its application to non-stationary conditions where  $t_{\rm acc} \lesssim t_{\rm ff}$. We propose that the effect of a density increase during the mass accumulation phase can lead to significant increases in the mass of fragments. This \emph{transport-driven super-Jeans fragmentation} is active in regions where the density is high, and the flow is turbulent, leading to the formation of massive fragments, which are precursors for massive stars.

Our new criterion can naturally explain why massive stars are formed in regions of high densities. We note that in the Jeans criterion, 
there is a negative correlation between the mass of the fragments and density, where $m_{\rm Jeans} \sim \rho^{-1/2}$. This negative correlation contradicts the fact that massive stars form exclusively in the densest parts of the interstellar medium \citep{2014prpl.conf..149T}. 
Our analysis can help to resolve this puzzle as the fragments formed in high-density, turbulent environments are more massive than previously thought, and is a possible mechanism which can explain the formation of massive stars in the universe.

Our modification should help to improve the numerical simulations of star-forming regions. The Jeans criterion has been a key part of the algorithm controlling the production of sink particles in numerical simulations. The new criterion should be placed in these algorithms to achieve consistent results when the conditions are dynamic.

\section{Transport-driven super-Jeans fragmentation}\label{sec2}

\subsection{The Jeans length}

We consider the fragmentation of a medium of density $\rho$, velocity dispersion $\sigma_{\rm v}$.
In this stationary case, the system contains two timescales, namely the free-fall time
\begin{equation}
    t_{\rm ff } = \frac{1}{(G \rho)^{1/2}}\;,
\end{equation}
and the sound crossing time
\begin{equation}
    t_{\rm cross } = \frac{l}{\sigma_{\rm v}}\;.
\end{equation}
The only dimensionless number is $\Pi_1 = t_{\rm ff} /t_{\rm cross} = \rm constant$. Using the Buckingham $\pi$ Theorem \citep{buckingham1914illustrations}, the governing equation of the system is $f({\Pi_1}) = 1$. Since there is only one variable $\Pi_1$, we conclude that in the     {\it Jeans fragmentation}:
\begin{equation}\label{eq:jeans}
 \Pi_1 = 1,\; t_{\rm  ff} = t_{\rm cross},\;  \lambda_{\rm Jeans} = c_{\rm s} / \sqrt{G \rho}. 
\end{equation}

The Jeans length $\lambda_{\rm Jeans} = \sigma_{\rm v} / \sqrt{G \rho}$ is the \emph{characteristic scale beyond which structures can collapse in a complex, self-gravitating system.}  This scale is relevant as long as a structure of density $\rho$ is supported by a pressure of the order $\rho \sigma_{\rm v}^2$. For example, the Jeans length is also the characteristic size beyond which
pressure-confined droplets become gravitationally unstable \citep{1955ZA.....37..217E,1956MNRAS.116..351B}.

We note that a more rigorous approach is to derive the Jeans length to add perturbations to a homogeneous system. In this case, the threshold wavelength appears naturally in the dispersion relation. Due to the highly idealized initial conditions,
this approach is rigorous in theory, but not particularly useful in realistic 
In this paper, the Jeans length is a threshold wavelength for gravitational stability, which is relevant for both cases.

\subsection{Non-stationary corrections to the Jeans length}\label{sec:setup}

% of a complex, multi-scale systems. A complex, turbulent flow can contain a set of structures of different properties, and only structures with $l > \lambda_{\rm Jeans}$ can collapse. It is through this line of through that the Jeans length is integrated in algorithms that controls the production of sink particles in numerical simulations \citep{2010ApJ...713..269F}.

In a dynamic environment, all parameters can change with time. We focus on the cases where the density changes, yet the velocity dispersion does not change significantly. This assumption is valid as long as the timescale of the temperature increase is longer than the free-fall time.

Our analysis aims to pin down the dominant wavelength of systems controlled complex, non-linear interactions between accretion and fragmentation. A positive $\dot \rho$ is the consequence of boundary conditions imposed on the system. Using mass conservations, 
\begin{equation}
    t_{\rm acc} = \rho / \dot \rho =  - \rho  (\nabla \cdot (\rho \vec{v}))^{-1} \;.    
\end{equation}
Examples of such density enhancements includes shocks in supersonic turbulence \citep{1998PhRvE..58.4501P,2011ApJ...730...40P}, density enhancements produced in converging flows \citep{1999A&A...351..309H,2007ApJ...657..870V} and high-density regions produced during cloud-cloud collisions  \citep{2021PASJ...73S...1F}. In the stationary case, the threshold wavelength $\lambda_{\rm Jeans}$ is selected through the interplay between pressure support and gravity. The very process of mass accumulation can increase the wavelength above which perturbations can grow. 

% The introduction of another timescale, the mass accumulation time $t_{\rm acc} = \rho / \dot \rho$ makes it no longer possible to derive the fragmentation length using simple dimensional arguments. 

% \subsection{Derivation}

To derive the fragmentation scales in this non-stationary setting, we start with the Buckingham $\pi$ Theorem \citep{buckingham1914illustrations}, and represent our system in several dimensionless parameters. These include 

\begin{equation}
    \Pi_1 = t_{\rm ff} / t_{\rm cross}
\end{equation}
 and  
 \begin{equation}
    \Pi_2 = t_{\rm ff} / t_{\rm acc} \;,
 \end{equation}
and in the    {\it super-Jeans fragmentation}, the governing equation should take the form 
\begin{equation}\label{eq:pi}
  \\ \Pi_1 = f(\Pi_2)\;, \lambda_{\rm Jeans, acc} = \lambda_{\rm Jeans} f(\Pi_2)\;.
\end{equation}

To determine the functional form of $f$, we introduce the asymptotic constraint that when the effect of density increase induced by accretion is negatable, the characteristic length for gravitational instability is the Jeans length. This means when  $t_{\rm acc} \rightarrow \inf $,  we expect
\begin{equation}
    \lim_{\Pi_2 \to 0}  f = 1 \;,
\end{equation} 
such that the fragmentation length is determined by Eq. \ref{eq:jeans}. The effect of accretion-induced density growth is significant when $\Pi_2 > 0$.

 When $\Pi_2 \gg 1 $, assuming that the system has no preferred scales, the relationship between the dimensionless variables should take the form of a power law, e.g. $\Pi_1 = \Pi_2^\alpha$. Taking these asymptotic constrains into considerations,  we have $f = (1 + f_0 \Pi_2^\alpha)$ or $f = (1 + f_0 \Pi_2)^\alpha$. 
 Since the behaviors of these two functional forms are nearly indistinguishable except at a smaller range of parameter space where $\Pi_2 \approx 1$ \footnote{These two forms may differ by 50\% when $\Pi_2 \approx 1$.}, we choose 
 \begin{equation}
    f = (1 + f_0 \Pi_2)^\alpha   
 \end{equation}
  for simplicity, where $f_0$ is a constant of the order of unity.
   Our remaining task is to determine the value of $\alpha$.

%  \subsection{General considerations}
%  Although the formalism we seek shares a very similar appearance to the Jeans
%  criterion, the underlying picture is different. In the Jeans criterion,
%  the fragmentation occurs in a homogeneous medium, and the analysis
%  is mainly concerned with the growth of density perturbations in the linear
%  regime.  In the non-stationary case, fragmentation occurs in dynamic environments
%  where density increase and fragmentation occurs simultaneously. 

%   In the
%  linear regime, the growth rate of all unstable wavelengths is bounded by the
%  free-fall time $t_{\rm ff} \approx 1 / \sqrt{G \rho}$, and then $t_{\rm acc} < t_{\rm ff}$, the growth of perturbations of wavelengths should be suppressed. However, assuming that some mechanism, such as turbulence, generates non-linear perturbations, the continuous mass inflow can further select the dominant wavelength based on how and at which rate they can accrete the newly-added gas. 

 % which makes it no longer possible to determine the behavior of the system through dimensional arguments alone. We have search for a similar-similar solution of the second kind \citep{1996sssi.book.....B},
% Following the classification by Barenblatt, this case is referred as \emph{self-similarity of the first kind}.   $f_0$ is a factor of the order of unity, and our task to derive the value of scaling exponent $\alpha$. 

%  {\bf Recall Jeans}

% , thus all wavelengths are effectively identical. Here, one key difference between these different fragmentation mode is the capacity to consume mass.

\subsection{Economy of arrangement -- Larger fragments accrete faster}
According to the  Jeans criterion, all wavelengths above $\lambda_{\rm Jeans}$ are unstable and can grow at the free-fall rate if pressure support is negligible.  Why does Nature select one configuration over the other? A key difference between these modes is their ability to accrete gas in the non-linear regime. 

We perform a thought experiment where we divide a medium into packets of different sizes and investigate the relation between fragmentation length and mass accretion. 
Assuming a fragmentation wavelength of $\lambda$, the mass of the fragments is $m \approx \rho \lambda^3$, and one can compute the accretion rate onto the fragments using $\dot m \approx G^2 m^2 \sigma_{\rm v}^{-3} \rho  $ \citep{1941MNRAS.101..227H,1952MNRAS.112..195B}, where accretion occurs inside a region of size $\lambda_{\rm acc} = G m \sigma_{\rm v}^{-2}$. The time for the fragments to increase their masses is
\begin{equation}
    t_{\rm acc,\, fragments} = m / \dot m = G \rho \lambda^{3} \sigma_{\rm v}^{-3}\;.
\end{equation}
For a medium filled with such packet of size $\lambda_{\rm acc}$, the maximum rate of density growth caused by accretion is characterized by the mass accumulation time
\begin{equation}
    t_{\rm acc}^* = \rho / \dot \rho = m / \dot m = t_{\rm double } =  G \rho \lambda^{3} \sigma_{\rm v}^{-3} \propto \lambda^{-3} \;.
\end{equation}

The fact that the mass accumulation time depends on $\lambda$  illustrates the key difference between these modes:  \emph{modes with larger $\lambda$ can have higher rates of density growth ($t_{\rm acc}^* \approx \lambda^{-3}$). }

% increase the density inside the accretion radius ($\lambda_{\rm acc}$)
%  varies drastically, where 
% modes with larger $\lambda$ having a higher capacity to consume mass,  supporting a higher rate of density increase.

% This result can also be understood imitatively, as gravity is a long-range force, and putting the centers of attraction together enables them to share the gravitational potential, leading to a higher rate of mass consumption. 
%  This difference is illustrated in Fig. 

% the mass accumulation time is $t_{\rm acc, \lambda} = \rho / \dot \rho = M \dot M = G^{-2}\rho^{-2} \lambda^{-3} \sigma_{\rm v}^3 $, where modes with large wavelengths have significant advantages in sustaining density growth. When the mass accumulation time $t_{\rm acc} = \rho / \dot \rho$ is a controlling parameter of the system, had chosen a fragmentation length is too small, the growth the density under self-gravity is too slow to accrete the newly-added material, and the growth of the perturbation is effectively suppressed. 

% accretion can be dominant if $t_{\rm ff} / t_{\rm acc}$ is large enough.

\subsection{Transport-driven super-Jeans fragmentation}
When an externally driven flow leads to a continuous increase in density, $t_{\rm acc} = \rho / \dot \rho$ becomes a control parameter of the system. We propose the \emph{hypothesis of transport-regulated fragmentation}: the fragmentation should in a way that to ensure that \emph{the rate of density increase caused by accretion should stay in sync with the density increase rate demanded by the boundary condition}, such that the fragmentation length can be determined by letting $t_{\rm acc} =t_{\rm acc}^*$. 

To understand why this condition is necessary, we consider a case where we choose a $\lambda$ that is too small. Even if such perturbations can grow, the newly added gas is too much for the system to consume.
 Only if we choose a $\lambda$ that is largely enough can the newly-added gas be accreted. \emph{A continuous injection of gas can suppress the growth of small-sized structures, favoring larger ones. }

Letting $t_{\rm acc} = \rho / \dot \rho =t_{\rm acc}^*$, we derive a critical wavelength of 
\begin{equation}
    \lambda_{\rm Jeans,\, acc-domiante} = t_{\rm ff} \sigma_{\rm v} (t_{\rm ff} / t_{\rm acc})^{1/3} = \lambda_{\rm Jeans} (t_{\rm ff} / t_{\rm acc})^{1/3}\;.
\end{equation}
Taking asymptotic constraints discussed in Sec. \ref{sec:setup} into account, the formula for  wavelength in the general case is
\begin{equation}
    \lambda_{\rm Jeans,\; aac} = \frac{\sigma_{\rm v}}{(G \rho)^{1/2}} (1 +  f_{\rm a}\, t_{\rm ff} / t_{\rm acc})^{1/3} = \lambda_{\rm Jeans} (1 + f_{\rm a}  \; t_{\rm ff} / t_{\rm acc})^{1/3} \;,
\end{equation}
where the characteristic mass is
\begin{equation}
    m_{\rm Jeans\, acc } =  \rho^{-1/2} G^{-3/2} \sigma_{\rm v}^3 (1 + f_{\rm a} \, t_{\rm ff} / t_{\rm acc}) = m_{\rm Jeans} (1 + f_{\rm a}  \; t_{\rm ff} / t_{\rm acc}) \;.
\end{equation}
where $\lambda_{\rm Jeans} \approx \sigma_{\rm v} / \sqrt{G \rho}$,  $m_{\rm Jeans} \approx \sigma_{\rm v}^3 G^{-3/2} \rho^{-1/2}$, 
$f_a \approx 1$ is a numerical prefactor of or unity, and its exact value can be derived by matching our formula with results from e.g. numerical simulations. 

The effect of accretion can lead to significant
increases in the characteristic mass of the fragments when the mass accumulation time
$t_{\rm acc}$ is short compared to the free-fall time.  Our approach of deriving the scaling exponent by demanding the conservation of some critical quantities (this quantity is the mass in our case) shares a similar spirit to how Kolmogorov derived the famous 5/3 energy spectrum of turbulence by demanding a constant energy flux between different scales \citep{1941DoSSR..30..301K}.

\begin{figure}
    \includegraphics[width = 0.5 \textwidth]{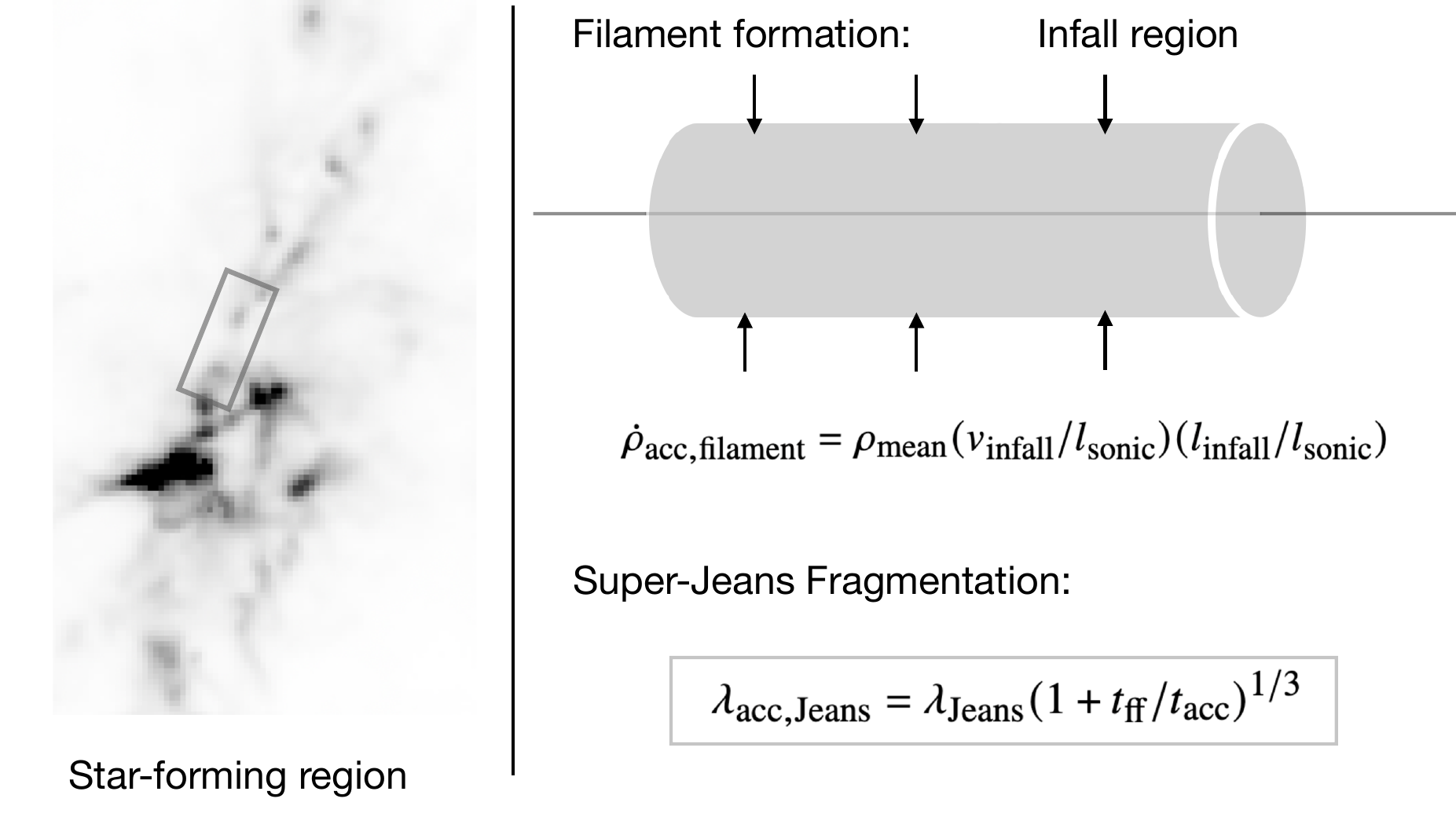}
    \caption{\label{fig:sketch} {\bf A sketch of transport-driven super-Jeans during filament assembly.} See Sec. \ref{sec:fila} for details.}
\end{figure}

\section{Application: Transport-driven super-Jeans fragmentation in massive star-forming regions} \label{sec:fila}
\subsection{Problem setup: Fragmentation during filament assembly} 
% To demonstrate the effect of accretion on fragmentation,
% 
In realistic settings, fragmentation rarely occurs in a homogeneous medium but can be bounded by various constraints (i.e. the size of the region). 
We consider the fragmentation occurring to a filament. The density of this filament is a function of time $\rho= \rho(t)$. The filament has a constant velocity dispersion $\sigma_{\rm v}$, and a width of $d \approx 0.1\;\rm pc$. The assumption that fragmentation occurs on filaments of $d\approx 0.1\;\rm pc$ is consistent with the results from Herschel observations towards nearby molecular clouds  \citep{2011A&A...529L...6A,2014prpl.conf...27A}, where the scale of 0.1 pc has been interpreted as sonic scale of the turbulence $l_{\rm sonic} $ \citep{2011A&A...529L...6A}. A sketch of this picture can by found in Fig. \ref{fig:sketch} 

One consequence of the time dependency is that the fragmentation scale and resulting mass are no longer constants. Due to the density increase, the fragmentation length decreases with time, and fragmentation occurs when $\lambda_{\rm Jeans, acc} \lesssim l_{\rm sonic}$.

% This assumption arises from the that both the Jeans criterion and our accretion-modified Jeans criterion require the medium to be homogeneous, yet in reality, 
%  the density structure of a cloud is highly inhomogeneous. These criteria are valid only if the critical wavelength is shorter than the coherence length of the gas (the scale at which the gas is approximately homogeneous). In reality, a reasonable choice of the coherence scale is the sonic length $l_{\rm sonic} \approx 0.1 \;\rm pc$. Inside the sonic length, the turbulent motion is subsonic such that the density fluctuations created by turbulence is no longer significant. Observations also indicate that dense cores, which result from gravitational fragments, stay on these filaments.   In our model, the density increases with time, leading to a steady decrease in the critical wavelength. Fragmenting only occurs when s smaller than the coherence length of the medium $l_{\rm sonic}$.  
 
%  This model is motivated by the fact that for fragmentation to occur, the  Motivated by the observational fact that the 
% molecular gas in molecular clouds is structured down to the scale of interstellar filaments, which have widths of around   in our model, as the fragmentation length decrease with time, we assume that  
% fragmentation should occur  at the time when $\lambda_{\rm fragmentation} \lesssim l_{\rm sonic} \approx 0.1 \;\rm pc$.

Should the fragmentation be Jeans-like, a unique critical density is determined through $\lambda_{\rm Jeans} = l_{\rm sonic}$, from which a single characteristic mass  of $\approx 0.9 \,M_{\odot}$ is implied (where $\rho_{\rm crit}$ can be determined via $\lambda_{\rm Jeans} = l_{\rm sonic}$, and the mass is $m_{\rm crit, Jeans} \approx \rho_{\rm crit} \lambda_{\rm Jeans}^3 $). The effect of accretion is to introduce another viable to the systems,  leading to a range of masses depending on how the gas density evolves.   

\begin{figure*}
    \includegraphics[width = 1.1 \textwidth]{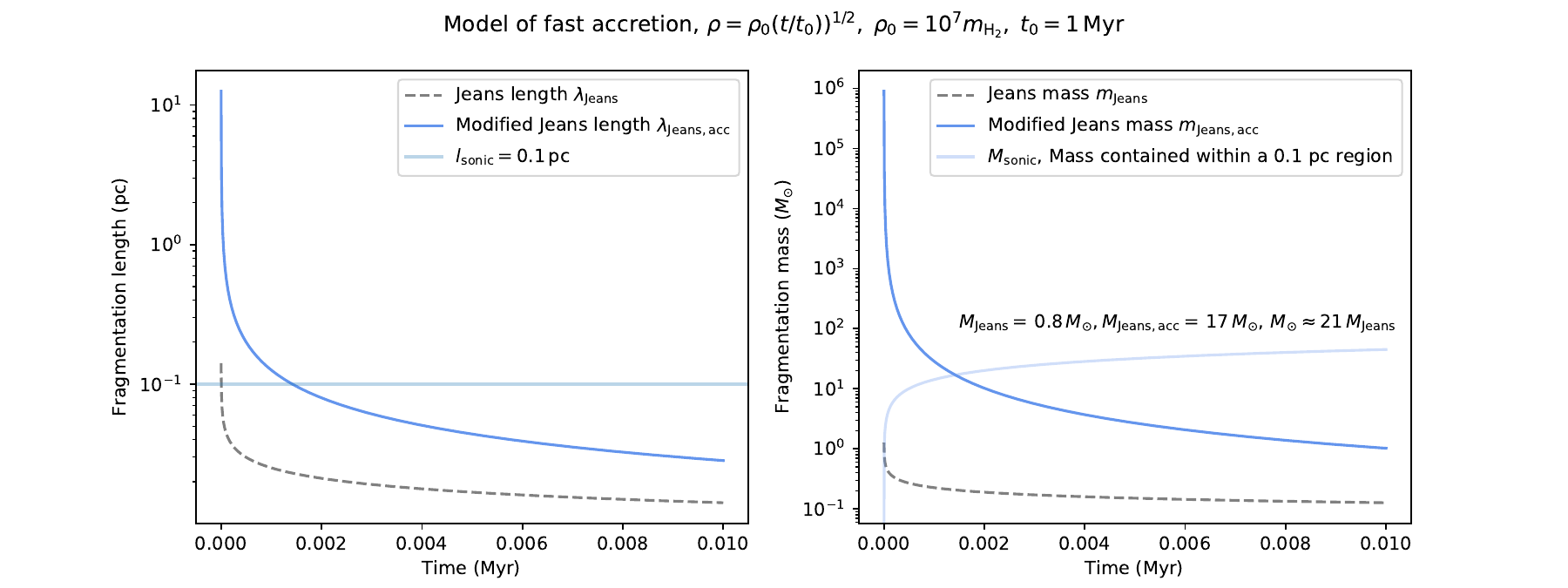}
    \caption{\label{fig:accretion} {\bf Effect of mass accumulation on fragmentation.} We consider a simple model where $\rho = \rho_0 (t/t_0)^{1/2}$, $\rho_0 = 10^7\, m_{\rm H_2}$, $t_0 =1\,\rm Myr$. {\bf Left Panel:} The Jean length $\lambda_{\rm Jeans}$ and the transport-modified Jeans length $\lambda_{\rm Jeans, acc}$ as a function of time. The horizontal line is the sonic scale $l_{\rm sonic} \approx 0.1 \;\rm pc$. Fragmentation starts when $\lambda_{\rm Jeans, acc} \approx l_{\rm sonic}$. {\bf Right Panel:} Jeans mass $m_{\rm Jeans}$ and transport-modified Jeans mass  $m_{\rm Jeans, acc}$ as a function of time. We also plot the $m_{\rm sonic}$, which is mass contained in regions of $l \approx l_{\rm sonic} \approx 0.1\;\rm pc$. Fragmentation occurs when $m_{\rm Jeans,\,acc} \lesssim m_{\rm sonic}$. We expect the fragments to have $m \approx 17 M_{\odot}$, which is much larger than the Jeans mass $\approx 21 \, m_{\rm Jeans}$.}
\end{figure*}

% To illustrate the effect of accretion, we consider a simple parameterized model where $\rho  = \rho_0  (t /t_0)
% )^{\gamma}$, and assume that  that fragmentation would occur when $\lambda_{\rm Jeans} \lesssim l_{\rm sonic} \approx 0.1 \;\rm pc $. We choose different values of $\rho_0/t_0$, and the evolution of the systems is plotted in Fig. \ref{fig:accretion}. When the density increase rate is large enough, e.g.  $\dot \rho =  10^7  {\rm cm}^{-3}\, m_{\rm H_2}\, {\rm Myr}^{-1} $, the effect of accretion becomes dominant, where fragments of 10 times the Jeans mass can be produced.  

To illustrate the effect of transport-driven mass accumulation, we consider a simple parameterized model where $\rho  = \rho_0  (t /t_0)
)^{\gamma}$, and assume that  that fragmentation would occur when $\lambda_{\rm Jeans} \lesssim l_{\rm sonic} \approx 0.1 \;\rm pc $. We choose $\gamma = 1/2$, where the growth is fast at the beginning and it slows down. This slow-down represents the gradual depletion of the gas reservoir. The evolution of a representative systems is plotted in Fig. \ref{fig:accretion}. When the density increase rate is large enough, e.g.  $\dot \rho =  10^7  {\rm cm}^{-3}\, m_{\rm H_2}\, {\rm Myr}^{-1} $, the effect of externally-driven flow become dominant, resulting in a mass that is one orders of magnitude larger than expected from the Jeans fragmentation.
% We expect our mechanism to be effective in Milky Way molecular clouds, as the accretion rate of  $\dot \rho =  10^7  {\rm cm}^{-3}\, m_{\rm H_2}\, {\rm Myr}^{-1} $, as required to form dense fragments, can be achieved in massive star-forming regions in the Milky Way. 

One can also derive characteristic mass analytically:
In our fiducial model where $\rho = \rho_0 (t/ t_0)^{1/2}$, assuming $l_{\rm sonic} = 0.1\;\rm pc$, $\sigma_{\rm v} = 0.2 \;\rm km \, s^{-1}$, $t_0 = 1\,\rm Myr$, the characteristic mass is (see \ref{sec:appen})
\begin{eqnarray}
    m 
    \approx 
     0.9  M_{\odot} + 1.7 M_{\odot} (\frac{\rho_0}{10^5\, m_{\rm H_2}})^{\frac{1}{2 }} \propto \rho_0^{1/2}\;,
    \end{eqnarray}
which is consistent with the numerical results.

\subsection{Super-Jeans fragmentation in massive star-forming regions}
In massive star-forming regions in the Milky Way, the mechanism of converging-flow-driven super-Jeans fragmentation can lead to the formation of fragments that are around one orders-of-magnitudes more massive than expected from the Jeans fragmentation. Recall that in our fiducial model where $\rho = \rho_0 (t/t_0)^{1/2}$, we require $\rho_0 = 10^7 {\rm cm}^{-3}\, m_{\rm H_2} $, and $t_0 = 1\;\rm Myr$ to produce a massive fragments of 17 $M_{\odot}$.

This density increase rate of $\dot \rho = 10^7  {\rm cm}^{-3}\, m_{\rm H_2}\, {\rm Myr}^{-1}$ is representative of the 
conditions in massive star-forming regions in the Milky Way. This is because the regions are dynamic, where the turbulent motion can create dense structures which relatively short time. 
 Assuming a simple cylindrical configuration where the gas is driven from the outside such that the filaments form at the center, we estimate a density increase rate of
\begin{equation}
    \dot \rho_{\rm acc, filament} = \rho_{\rm mean} (v_{\rm infall} / l_{\rm sonic}) (l_{\rm infall} / l_{\rm sonic}) \;,
\end{equation}
where $\rho_{\rm mean}$ is the mean density of the medium, $v_{\rm infall}$ in an infall speed, $l_{\rm sonic}$ is the sonic scale, and $l_{\rm infall}$ is the scale at which infall is initiated. Toward a typical  
high-mass star-forming region such as the IRDC 18223 \citep{2015A&A...584A..67B}, we adopt $\rho_{\rm mean} = 10^5\, {\rm cm}^{-3}$, which is the measured density of the region, and $v_{\rm infall} = 2.5\;\rm km\,s^{-1}$, which is the measured velocity dispersion of the region. We further assume
$l_{\rm sonic} = 0.1\;\rm pc$, and $l_{\rm infall} = 0.5 \;\rm pc$, which is a fraction of the size of the region. From these, we derive a density increase rate of $\approx 1 \times 10^7\;\rm cm^{-3}\, {\rm Myr}^{-1}$, consistent with what is required in our fiducial model. 

This transport-driven super-Jeans fragmentation can explain the puzzling fact why the separation between dense cores in massive star-forming regions is larger, and the cores are more massive than expected from the Jeans fragmentation \citep{2009ApJ...696..268Z,2014MNRAS.439.3275W,2018A&A...616L..10F,2023MNRAS.520.3259X} as revealed by ALMA observations. However, other studies have concluded that the separations between dense cores are consistent with thermal Jeans fragmentation \citep{2017ApJ...849...25L,2019ApJ...886..102S,2019ApJ...886...36S,2020ApJ...894L..14L}. These conflicting conclusions can result from the intrinsic difference in mass accumulation rate between different regions, the differences in techniques used, or the failure to account for the further evolution of core separation after fragmentation. Nevertheless, one can conclude that super-Jeans separations are observed at least in some regions, where transport-driven super-Jeans fragmentation is a promising explanation. Once formed, these super-Jeans fragments can suppress the subsequent fragmentation through tidal forces, as a recent study has demonstrated  \citep{2023MNRAS.tmpL.151L}.

\section{Conclusions}
The formula proposed by Jeans predicts the characteristic mass of fragments produced in gravitational fragmentation. A limitation of the Jeans criteria is that the setting is stationary, as in reality, the density is subject to a constant change in dynamic environments such as star-forming regions. This time-dependency would become a significant issue when the mass accumulation time $t_{\rm acc} = \rho / \dot \rho =  - \rho  (\nabla\cdot  (\rho \vec{v}))^{-1}$ becomes shorter than the free-fall time $t_{\rm ff} = 1 / \sqrt{G \rho}$, where we expect the behavior of the system to be different.  

When gas is supplied to a region from the outside, the growth of larger perturbations is preferred over smaller ones,  because larger fragmentation modes can consume mass at higher rates.  In this non-stationary gas, the Jeans length becomes 
\begin{equation}
    \lambda_{\rm acc, Jeans} = \lambda_{\rm Jeans} (1 + t_{\rm ff} / t_{\rm acc})^{1/3} \;,
\end{equation}
and the Jeans mass becomes
 \begin{equation}
    m_{\rm acc, Jeans}  = m_{\rm Jeans} (1 + t_{\rm ff} / t_{\rm acc})\;,
 \end{equation}
 where the term $t_{\rm ff} / t_{\rm acc}$ represents the effect of density increase caused by an externally driven flow. 
 Under a realistic mass inflow rate as estimated from a typical high-mass star formation region, we expect $m\approx 17\, M_{\odot}$, which is one order of magnitude larger than the Jeans case. This transport-driven super-Jeans fragmentation is one key mechanism leading to the formation of massive stars in galaxies. 

 Transport-driven super-jeans fragmentation is thus a key mechanism for massive star formation. In contrast to the earlier proposals where the role of turbulence is to regulate the star formation process \cite{2005ApJ...630..250K}, recent developed tend to favor a link between a dynamical, active environment with the formation of more massive objects \citep{2019MNRAS.490.3061V,2020ApJ...900...82P}.  The significant increase of the Jean mass under such a dynamical environment we propose is along the same direction.   We expect that future applications of our results to these dynamical settings can lead to a better understanding of the formation of massive stars in the universe.

 \section*{Acknowledgements}
 
We would like to thank the referee for careful readings of the paper and for constructive comments. GXL acknowledges support from NSFC grant No. 12273032 and
12033005. This work is motivated from a collaboration with Feng-Wei Xu (Peking University) and colleagues from the ATOMS collaboration.
Guang-Xing Li would like to thank Prof. Andreas Burkert for sharing his curiosity and excitement about equations. For Prof. Xun Shi for her patience with the timescales, and thank friends in the cycling group, Prof. Chandra B. Singh, Qiqi Jiang for reviving his interest in equations. 

%%%%%%%%%%%%%%%%%%%%%%%%%%%%%%%%%%%%%%%%%%%%%%%%%%%%%%%
%%% Appendix sections. ??????, ????
%%%%%%%%%%%%%%%%%%%%%%%%%%%%%%%%%%%%%%%%%%%%%%%%%%%%%%%
\section*{Data availability statement}
No proprietary data was used during the preparation of the manuscript.

%\section{Name}

%\end{appendix}

%\begin{appendices}
%\section{Appendix}
%\end{appendices}
%\appendix
\bibliographystyle{mnras}

\bibliography{paper}

%\appendix

\appendix
\section{Characteristic mass}\label{sec:appen}
We consider a model of density growth where
\begin{equation}
    \rho = \rho_0 (t / t_0)^{\gamma}\;.
\end{equation}
Assuming that  fragmentation occurs on filaments of the width of $d\approx l_{\rm sonic} \approx 0.1 \,\rm pc$, we study the effect of accretion on the evolution of such a system. The free-fall time is
\begin{equation}
    t_{\rm ff} = 1 / \sqrt{G \rho} = 1 / \sqrt{G \rho_0}\, (t / t_0)^{-\gamma / 2}\;,
\end{equation}
and the mass accumulation time is
\begin{equation}
   t_{\rm acc} = \rho / \dot \rho = \gamma t\;.
\end{equation}

The increasing density as assumed in the model implies a decreasing fragmentation length. Fragmentation occurs when $\lambda \lesssim d \approx l_{\rm sonic} \approx 0.1\;\rm pc$. In the case of the Jeans fragments, the critical density can be solved via $\lambda_{\rm Jeans} = l_{\rm sonic}$, where

\begin{equation}
    \rho_{\rm crit, Jeans} = \frac{\sigma_{\rm v}^2}{l_{\rm sonic}^2 G}\;,
\end{equation}
and a mass of 
\begin{equation}
 m_{\rm crit, \,Jeans} =   \rho_{\rm crit,\, Jeans} l_{\rm sonic}^3 = \frac{l_{\rm sonic}\sigma_{\rm v}^2}{ G} \approx 0.9 \,m_{\odot}\;,
  \end{equation}    
can be derived. 

  Next we derive the characteristic mass in the accretion-dominated regime. When $t_{\rm ff} \gg t_{\rm acc}$ The accretion-modified Jeans length 
is 
\begin{equation}
    \lambda_{\rm Jeans, acc-dominated} = \frac{\sigma_{\rm v}}{(G \rho)^{1/2}} (t_{\rm ff} / t_{\rm acc})^{1/3}\;,
\end{equation}
from which we can derive the time at which fragmentation occurs by solving $\lambda_{\rm Jeans, acc-dominatated} = l_{\rm sonic}$:
\begin{equation}
    \Big{(} \frac{t_{\rm fragmentation}}{t_0} \Big{)}^{-2 \gamma -1 } =   \gamma t_0 (\frac{l_{\rm sonic}}{\sigma_{\rm v}})^{3} (G \rho_0)^2  \;,  
\end{equation}
where the mass of the fragments is 
\begin{eqnarray}
   m_{\rm Jeans, acc-dominated} &=& \rho(t_{\rm fragmentation})  l_{\rm sonic}^3 \\
   &=& \rho_0 \big{(}  \gamma t_0 \big{(}\frac{l_{\rm sonic}}{\sigma_{\rm v}} \big{)}^3 (G \rho_0)^2  \big{ ) }^{\frac{- \gamma}{2 \gamma + 1}} l_{\rm sonic}^3 \\
   &=& G^{\frac{-2 \gamma}{2 \gamma + 1}} \rho_0^{\frac{1}{2 \gamma + 1}} l_{\rm sonic}^{\frac{3 \gamma + 3}{2 \gamma + 1}} \sigma_{\rm v}^{\frac{3 \gamma}{2 \gamma + 1}} t_0^{\frac{-\gamma}{2 \gamma + 1}} \gamma^{\frac{- \gamma}{2 \gamma + 1}} \;.
\end{eqnarray}

Consider a fiducial case where $\gamma=1/2$, we have 
\begin{eqnarray}
   \frac{ m_{\rm Jeans, acc-dominated}}{1.7 M_{\odot} } 
   \approx 
    (\frac{\rho_0}{10^5 m_{\rm H_2}})^{\frac{1}{2 }} (\frac{l_{\rm sonic}}{0.1 \;\rm pc})^{\frac{9}{4}} (\frac{\sigma_{\rm v}}{0.2\,\rm km\;s^{-1}})^{\frac{3 }{4}} (\frac{t_0}{1\,\rm Myr})^{\frac{-1}{4}}\;.
\end{eqnarray}

%\end{appendices}

%%%%%%%%%%%%%%%%%%%%%%%%%%%%%%%%%%%%%%%%%%%%%%%%%%

%%%%%%%%%%%%%%%%% APPENDICES %%%%%%%%%%%%%%%%%%%%%

% \appendix
% \section{Journal abbreviations}

%%%%%%%%%%%%%%%%%%%%%%%%%%%%%%%%%%%%%%%%%%%%%%%%%%

% Don't change these lines
\bsp	% typesetting comment
\label{lastpage}
\end{document}